\documentclass[twocolumn]{article}
\usepackage[a4paper]{geometry}
\usepackage[utf8]{inputenc}
\usepackage{authblk}
\usepackage{natbib}
\usepackage{graphicx}
\usepackage{xcolor}  
\usepackage{anysize}
\usepackage{wrapfig} 
\usepackage{caption}
\usepackage{amssymb}
\usepackage{float}
\usepackage{abstract}
\usepackage{doi}

\usepackage[T1]{fontenc}
\usepackage{authblk}

\marginsize{2.5cm}{2.5cm}{2.5cm}{2.5cm}

\usepackage[compact]{titlesec}
\titlespacing{\section}{0pt}{2ex}{1ex}
\titlespacing{\subsection}{0pt}{1ex}{0ex}
\titlespacing{\subsubsection}{0pt}{0.5ex}{0ex}

\author{A. Janiuk$^{1}$\thanks{agnes@cft.edu.pl}, B. James$^{1}$, K. Sapountzis$^{1}$
\\
      $^{1}${Center for Theoretical Physics, Polish Academy of Sciences, Al. Lotnik\'ow 32/46, 02-668 Warsaw, Poland}
}
\vspace{-20cm}

\title{Cosmic Gamma Ray Bursts}
\date{Key words: accretion, accretion disks – black hole physics – Magnetohydrodynamics (MHD)}

\begin{document}

\twocolumn[
  \maketitle
  \begin{onecolabstract}

    Gamma ray bursts (GRBs) are astronomical phenomena detected at highest energies. The gamma ray photons carry energies on the order of mega-electronovolts and arrive to us from the point-like sources that are uniformly distributed on the sky. A typical burst has a form of a pulse that lasts for about a minute.  As the Earth’s atmosphere is not transparent to the very high energy radiation, the bursts are detected by means of the telescopes onboard satellites that are placed on the orbit \citep{swift2004}.
    The total energetics of GRB events, which is given by the integrated energy flux by the detector unit area, implies that we are witnessing very powerful explosions, where an enormously great power is released within a short time. There is only one way to obtain such huge energies in cosmos: the disruption of a star \citep{Paczynski1986}.
    
 \end{onecolabstract}
]


\section{Introduction}

Multi-wavelength and multi-messenger observations show that black holes are the central
engines responsible of the most violent astrophysical events such as, for instance, active nuclei of galaxies, X-ray binaries, core-collapse supernovae or gamma-ray bursts. This central engine is subject to strong gravity, strong electromagnetic fields and rotation. The governing physical laws of such engines are well known (General Relativistic MagnetoHydroDynamics, hereafter GRMHD) but are nonlinear, time-dependent and multidimensional. Thereby, it is necessary to
develop a numerical approach to simulate their evolution and observational appearance where a first-principles theory can not be achieved.

In order to produce a gamma ray burst, the black hole is launching a jet of plasma. In these jets, plasma is expanding with ultra-relativistic velocities. Particles are accelerated in the strong magnetic field and produce high energy radiation.
The jet is powered by the Blandford-Znajek mechanism which can extract energy from a rotating black hole \cite{blandford1977}. This process requires a magnetized accretion disk with a strong poloidal magnetic field generated around a spinning black hole.
The gamma-ray emission produced in the jet at large distances is not uniform and short time-scale variability suggests that it originates close to the black hole.

\section{Two classes of gamma ray bursts}

Two disctinct classes of Gamma Ray bursts have been identified and are known to constitute statistically distinct populations of sources \citep{crissa1993}.
The long bursts cluster around few tens of seconds of gamma ray emission, while the short bursts are typically lasting a fraction of a second only. Also, their characteristic spectral energy distributions have distinct characteristics (see Table \ref{table1}).

\begin{table}[]
  \Huge
	\centering
	\resizebox{0.48\textwidth}{!}{%
		\begin{tabular}{|c|c|c|c|}
			\hline
			Type  & Mean Duration & Peak in Energy Spectrum    & Origin                      \\ \hline
			Long  & 25 [s]          & $\log (E_{\rm peak}) = 2.2$ [keV] & Collapse of massive star    \\ \hline
			Short & 0.7 [s]         & $\log (E_{\rm peak}) = 2.7$ [keV] & Merger of two compact stars \\ \hline
		\end{tabular}%
	}
	\caption{Properties of short and long GRBs}
	\label{table1}
\end{table}

\subsection{Long gamma ray bursts}

Already from 1990’s we know that the GRBs originate mostly in the distant galaxies, and many of them are associated with supernovae. They have to be in fact special supernova types (only 10 per cent of them meet the criteria), because the core of collapsing star needs to form a black hole, surrounded by a disk composed from the remnant matter from the stellar envelope. It is the accretion of matter onto a rotating black hole that is able to provide energy large enough to account for the observed properties of GRB phenomena. This process is relatively long (several tens – hundreds seconds).

\subsection{Short gamma ray bursts}

Another mechanism of producing a shorter GRB is the coalescence of two neutron stars. A transient structure is then formed, and collapses to a black hole. The surrounding remnants of dense matter form a disk composed of elementary particles and neutrinos. The process of disk accretion, mediated by magnetic fields provides power  to extract the rotational energy of the black hole and launch a relativistic jet \citep{janiuk2010}.
This collimated outflow is where the gamma rays are produced.

\section{Central engine}

\subsection{Progenitors}

\begin{figure}[H]
\centering
\includegraphics[width=0.4\textwidth]{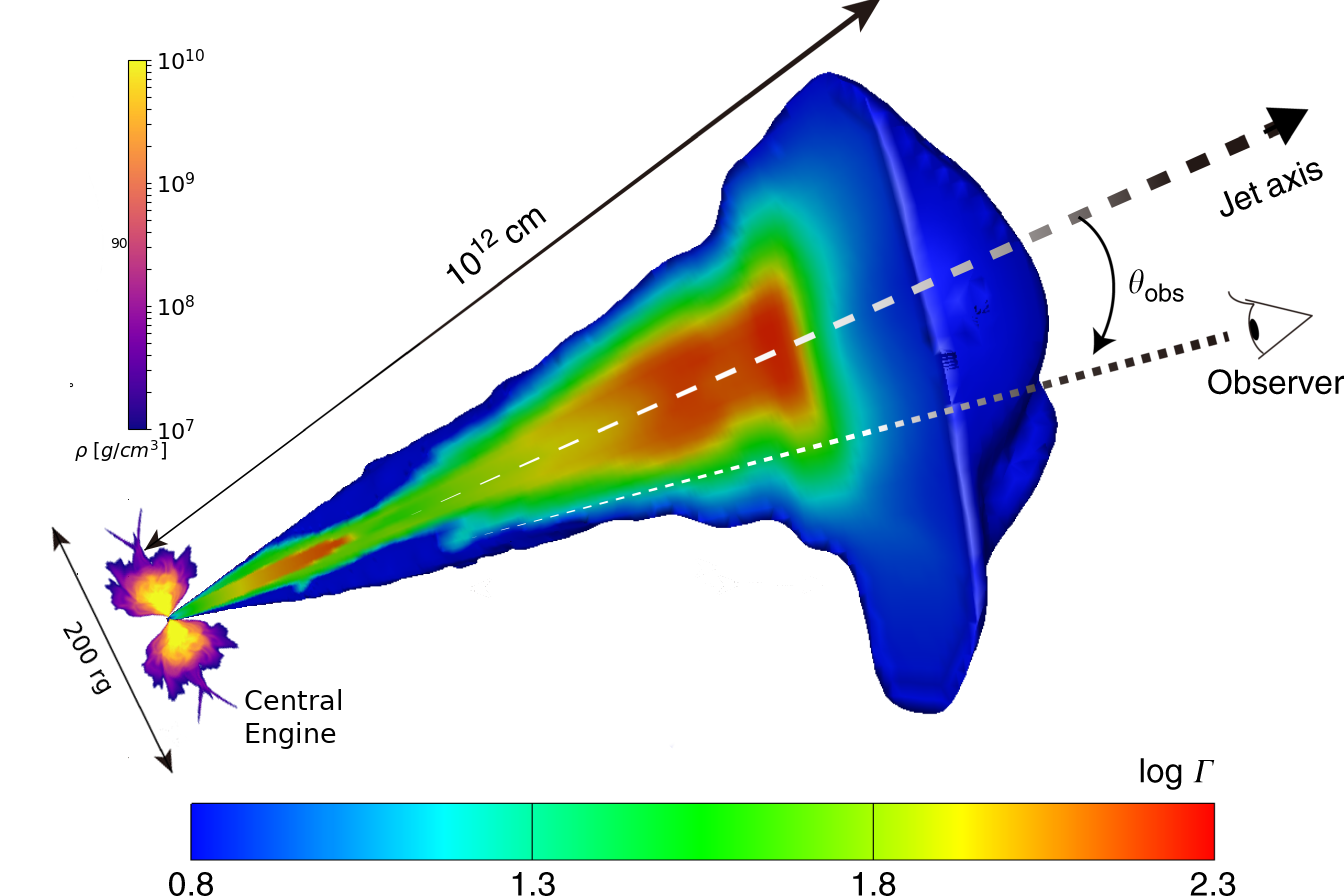}
\caption{Schematic idea of GRB. The ralativistic jet is ejected (Lorentz factor about 100). Its origin is the “central engine” where the black hole sits. The jet emits gamma ray radiation, collimated towards the observer.}
\label{fig:schematic}
\end{figure}

The collapsar scenario is able explain the long-duration GRBs, while the short GRBs are associated with the mergers of compact objects. In the long GRB case, the energetics of the explosion is consistent with the gravitational mass of the progenitor star: $E=GM_{\star}^{2}/R \approx 10^{54}$ [erg]. The jet has to have a highly relativistc speed, with a bulk lorentz factor of $\Gamma\sim 100$, in order to be able to break through the stellar envelope.
Also, the duration, time variability, and the following afterglow emission of these GRBs at lower energies are consistent with the collapsing massive stars that reside in star-forming host galaxies.
In the short GRB case, the compact binary megergers are almost equally energetic, while the duation of the burst is set by the viscous timescale of an accretion disk formed from the tidally-disrupted remnant,
$t_{vis}\approx \alpha_{\rm vis}^{-1} (R_{\rm disk}^{3}/GM_{NS})^{1/2} (H/R_{\rm disk})^{-2}$, and is about 0.5 [s] for typical parameters of such system.
The highly relativistic speeds of the jets are also supported by the observation of non-thermal energy spectra, which otherwise would not be possible because
of a lare optical depht due to the electron-positon pair production (see \cite{Thompson1994}). 
Recently, the discovery of the binary neutron star (NS-NS) mergers in the gravitational wave observation made by LIGO (GW170817 and GW190425), as well as the detection of associated electromagnetic counterparts, provided a direct proof of the NS-NS system being sources of short GRBs (see review by \cite{Janiuk2018}).
Schematic view of a central engine of a GRB, in a unified aproach, is depicted in Fig. \ref{fig:schematic}.

\subsection{Numerical modeling of the engine and jet structure}

We compute the structure and evolution of black hole accretion disks using the numerical simulations, governed by the equations of general relativistic magnetohydrodynamics (GRMHD).
In particular, such disks and outflows can be found at the base of
relativistic jets in the extragalactic sources, like blazars,
or gamma ray bursts.
Long-lasting, detailed computations are essential to properly determine the
physics of these jets, and confront the theoretical models with observables available from astrophysical observatories in space and from ground-based detectors.

Our numerical scheme works 
in a conservative manner, by solving a set of non-linear equations at each time-step, to advance the conserved quantities from one time step to the next.
The efficiency of computations is enhanced thanks to
the code parallelisation.
We use the Message Passing Interface (MPI) techniques to distribute the
computational grid over the threads.
Such calculations are computationally demanding, and also required special
fine-tuning of the code algorithm.

Our calculations are started with
a steady state model of the flow,
based on the analytical, equilibrium solution driven by the main physical parameters of the black hole accretion disk (namely, BH mass, its spin, and size and density of the accreting torus).
Properties of the pressure equilibrium torus around a black hole, supported by
a constant specific angular momentum over radius, were found by
\cite{Moncrief1976} and \cite{Chakrabarti1985}.
In the latter model, the angular momentum distribution is chosen to have a power law distribution, which differs from the default disk model of \cite{Moncrief1976}, where the angular momentum is assumed to be constant in the disk. This model will allow us to create a initial torus with a large amount of poloidal magnetic flux.

Starting from the Chakrabarti solution, implemented as the initial condition for
our simulations, we use a dynamical scheme, and we
follow the flow evolution. This is achieved by solving numerically the continuity and momentum-energy conservation equations in GR MHD framework:

\begin{equation}
(\rho u_{\mu})_{;\nu} = 0; 
\hspace{1cm}
T^{\mu\nu} = T^{\mu\nu}_{gas}+T^{\mu\nu}_{EM} = 0
\end{equation}

\begin{equation}
  T^{\mu\nu}_{gas} = \rho h u^{\mu} u^{\nu}+pg^{\mu\nu} = ( \rho + u + p)u^{\mu} u^{\nu}+pg^{\mu\nu}
  \end{equation}
  \begin{equation}
    T^{\mu\nu}_{EM} = b^{2} u^{\mu} u^{\nu}+\frac{1}{2} b^{2} g^{\mu\nu}- b^{\mu} b^{\nu};  b^{\mu}=u^{*}_{\nu}F^{\mu\nu}
  \end{equation}
  
Here we account for both
the gas and electromagnetic components of the stress tensor, 
$u^{\mu}$ is the four-velocity of gas, $u$ is internal energy, $\rho$ is density, $p$ denotes pressure, and  $b^{\mu}$ is the magnetic four-vector.
Here, $F$ is the Faraday tensor, and in the force-free approximation the Lorentz force vanishes,
$E_{\nu}=  u^{\nu}F^{\mu\nu} = 0$.

The model system of equations is supplemented with an equation of state, in the polytropic form\begin{equation}
p_{\rm g} = K \rho^{\gamma}
\end{equation}
where $p_{\rm g}$ is the gas pressure, $\rho$ is density, $K$ is constant specific entropy, and
the polytropic index $\gamma=4/3$ is typically used in the context of gamma ray bursts.
The plasma $\beta-$parameter is defined
as the ratio of the fluid's thermal to the magnetic pressure, $\beta \equiv
p_{\rm g} / p_{\rm mag}$.  We normalize the magnetic field in the torus to
have $\beta=(\gamma - 1) u_{\rm max}/( b^{2}_{\rm max}/2) $, where $u_{max}$ is the
internal energy at the torus pressure maximum radius, $r_{\rm max}$. 

\begin{figure}[H]
\centering
\includegraphics[width=0.4\textwidth]{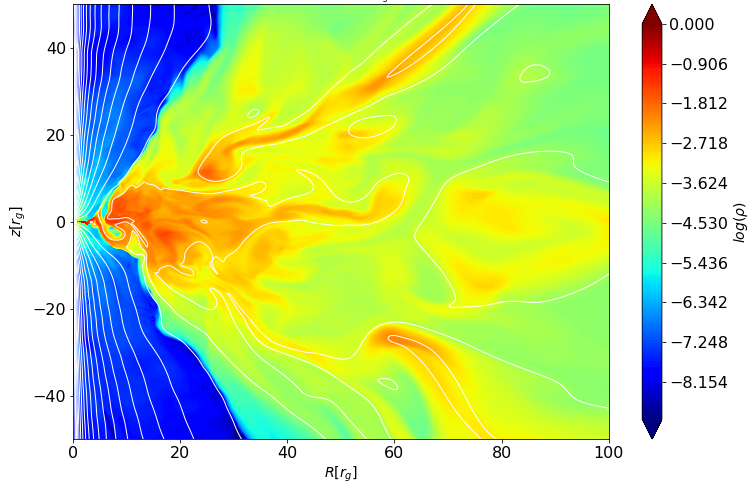}
\includegraphics[width=0.4\textwidth]{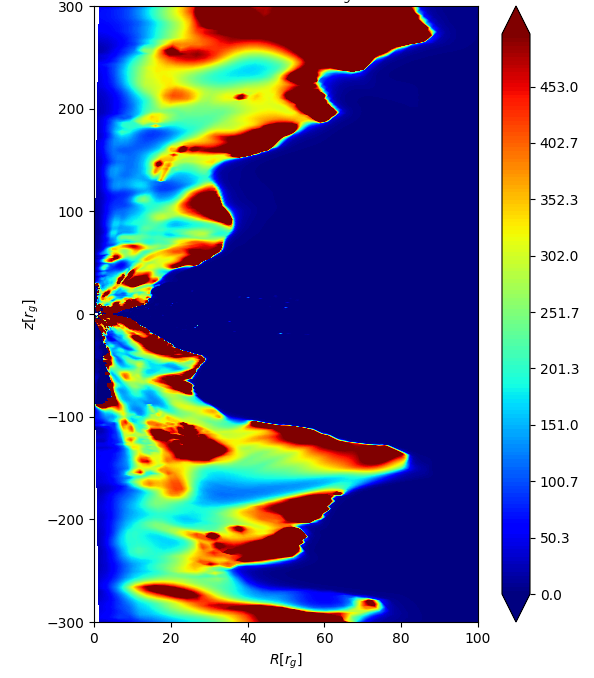}
\caption{{\bf Top:} Distribution of mass density and configuration of magnetic field lines in the region close to the black hole, at the base of gamma ray burst. 
{\bf Bottom:} Structure of jet launched from rotating black hole, in terms of its total energetics (thermal plus Poynting energy). The results from axisymmetric numerical simulation are shown, for black hole spin $a=J/M^{2}$ = 0.7. The snapshot is taken at time $t=2000 ~[t_{\rm g}]$. 
}
\label{fig:engine}
\end{figure}

Resulting evolved structure of the GRB central engine (disk in the upper panel, and jet in the bottom panel) is shown in Figure \ref{fig:engine}. The snapshot is taken at time $t=2000 ~[t_{\rm g}]$, where $t_{\rm g} = GM/c^{3}$ denotes geometric time unit, and scales with black hole mass. For GRB engine, typical mass of central black hole is in the range from $M \sim 3$ up to $M \sim 30$ Solar masses.
Note that in the evolved state we are able to obtain the physically motivated structure of the accretion torus in the equatorial plane of the rotating black hole, and low-density polar funnels, with dominant magnetic field.
These funnels carry both thermal and Poynting energy along the jet axis.
The ultimate bulk Lorentz factor in the jet, achieved at 'infinity' (outside the computational domain) will possibly reach the order of magnitude of this total energetics parameter, so $\Gamma \sim$ few hundreds.

+

\section{Discussion}


\begin{figure}[H]
\centering
\includegraphics[width=0.4\textwidth]{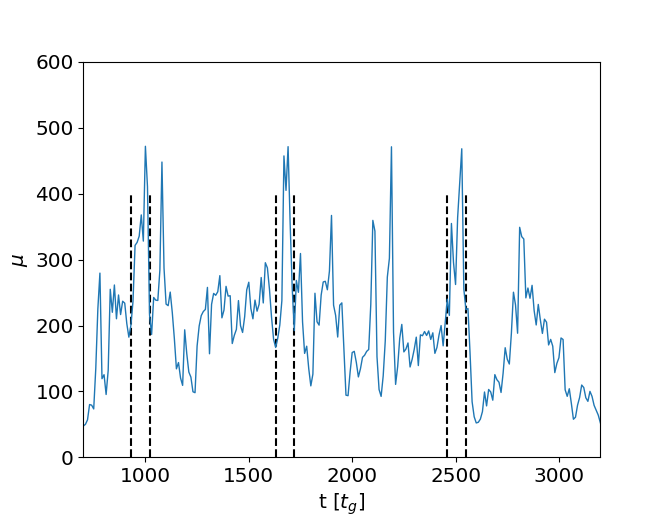}
\caption{Time variability of the jet in terms of its total energetics (thermal plus Poynting energy), as measured during the simulation at a chosen point in the jet (located at an angle $\theta=5^{\circ}$ from the jet axis). The result of numerical 2-dimensional simulation is shown, for black hole spin $a=J/M^{2}=0.7$, and magnetic field normalized in torus with $\beta(r_{\rm max})=60$.
The dashed lines show the characteristic timescale of the MRI instability, measured as the maximum growth rate timescale (Gammie 2004) which correlates with the pulse width.}
\label{fig:variability}
\end{figure}

\begin{figure}[H]
\centering
\includegraphics[width=0.4\textwidth]{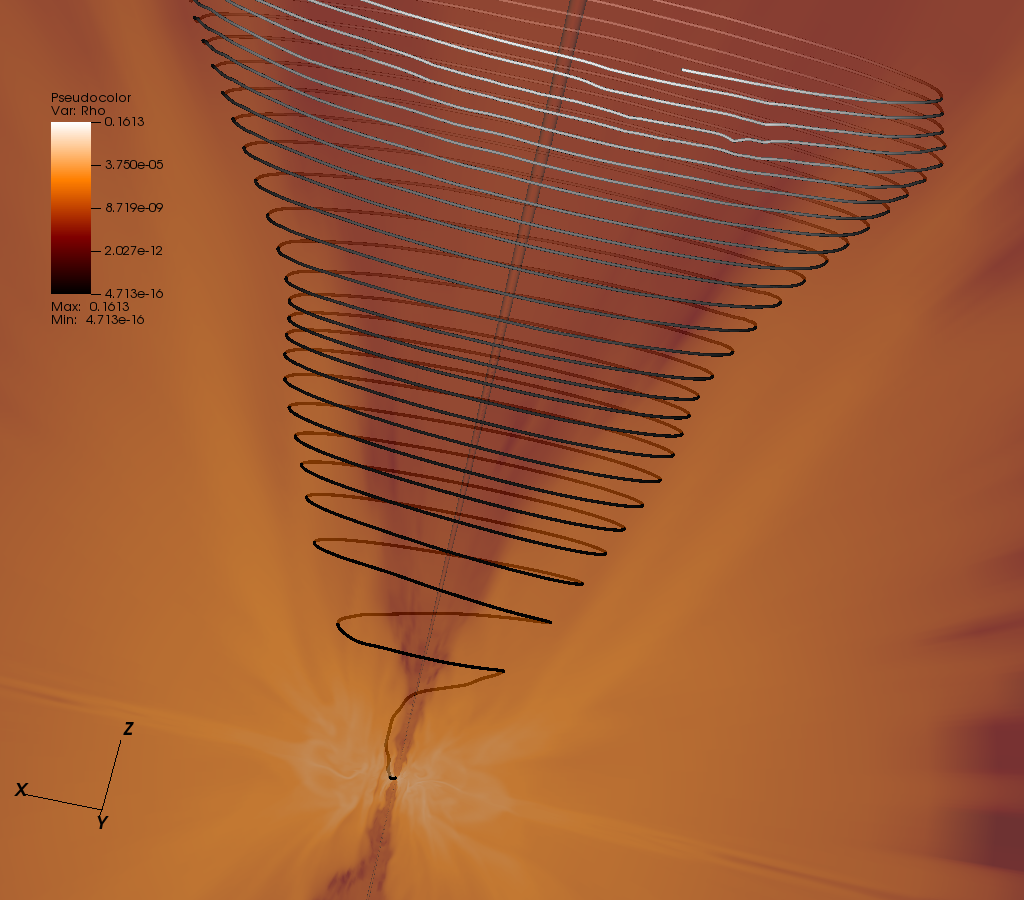}
\caption{Structure of the accretion flow and large scale magnetic field in the jet. Figure shows the result of 3-dimensional numerical simulation. Exemplary field line is depicted as launched from the black hole horizon, with dominant toroidal field component.}
\label{fig:jet3D}
\end{figure}

Our numerical simulations, performed by the computational astrophysics group at the Center for Theoretical physics, PAS, provide working tools to model the central engine of both long and short GRBs.
The unified approach is possible because we scale the size and timescale of the system by converting the geometric units to physical ones, as adequate in the GR MHD modeling.
We also notice that the variability timescales, modeled in our simulations as the time variability of the jet energetics (see Fig. \ref{fig:variability}), are 
governed by the timescale of the magneto-rotational instability in the accretion disk.
In order to properly govern the regime of long-duration GRB, with a sustained magnetic turbulence in the engine, full 3-dimensional simulations are needed. This is the work in progress carried by our group (Fig. \ref{fig:jet3D}).

Such models, with a good spatial resolution to solve physical process close to the black hole, need large computational resources. We ran already
several series of calculations on the OKEANOS supercomputer of the Warsaw ICM supercomputing center, and in the Prometheus supercomputer within the
PL-Grid. The purpose is to study the polarimetric signatures of strong magnetic fields near event horizon of a black hole in the Galaxy center  (Moscibrodzka et al., 2020, in preparation).


\section*{Acknowledgement}
 
The authors acknowledge the financial support by the grant No. 2016/23/B/ST9/03114 and 2019/35/B/ST9/04000 from the Polish National Science Center. We have used the computational resources of the Warsaw ICM through grant Gb79-9, and the PL-Grid through the grant grb3.

\bibliographystyle{unsrtnat}
\bibliography{references}

\begin{thebibliography}{9}
\providecommand{\natexlab}[1]{#1}
\providecommand{\url}[1]{\texttt{#1}}
\expandafter\ifx\csname urlstyle\endcsname\relax
  \providecommand{\doi}[1]{doi: #1}\else
  \providecommand{\doi}{doi: \begingroup \urlstyle{rm}\Url}\fi

\bibitem[{Gehrels} et~al.(2004){Gehrels}, {Chincarini}, {Giommi}, {Mason},
  {Nousek}, {Wells}, {White}, {Barthelmy}, {Burrows}, {Cominsky}, {Hurley},
  {Marshall}, {M{\'e}sz{\'a}ros}, {Roming}, {Angelini}, {Barbier}, {Belloni},
  {Campana}, {Caraveo}, {Chester}, {Citterio}, {Cline}, {Cropper}, {Cummings},
  {Dean}, {Feigelson}, {Fenimore}, {Frail}, {Fruchter}, {Garmire}, {Gendreau},
  {Ghisellini}, {Greiner}, {Hill}, {Hunsberger}, {Krimm}, {Kulkarni}, {Kumar},
  {Lebrun}, {Lloyd-Ronning}, {Markwardt}, {Mattson}, {Mushotzky}, {Norris},
  {Osborne}, {Paczynski}, {Palmer}, {Park}, {Parsons}, {Paul}, {Rees},
  {Reynolds}, {Rhoads}, {Sasseen}, {Schaefer}, {Short}, {Smale}, {Smith},
  {Stella}, {Tagliaferri}, {Takahashi}, {Tashiro}, {Townsley}, {Tueller},
  {Turner}, {Vietri}, {Voges}, {Ward}, {Willingale}, {Zerbi}, and
  {Zhang}]{swift2004}
N.~{Gehrels}, G.~{Chincarini}, P.~{Giommi}, K.~O. {Mason}, J.~A. {Nousek},
  A.~A. {Wells}, N.~E. {White}, S.~D. {Barthelmy}, D.~N. {Burrows}, L.~R.
  {Cominsky}, K.~C. {Hurley}, F.~E. {Marshall}, P.~{M{\'e}sz{\'a}ros}, P.~W.~A.
  {Roming}, L.~{Angelini}, L.~M. {Barbier}, T.~{Belloni}, S.~{Campana}, P.~A.
  {Caraveo}, M.~M. {Chester}, O.~{Citterio}, T.~L. {Cline}, M.~S. {Cropper},
  J.~R. {Cummings}, A.~J. {Dean}, E.~D. {Feigelson}, E.~E. {Fenimore}, D.~A.
  {Frail}, A.~S. {Fruchter}, G.~P. {Garmire}, K.~{Gendreau}, G.~{Ghisellini},
  J.~{Greiner}, J.~E. {Hill}, S.~D. {Hunsberger}, H.~A. {Krimm}, S.~R.
  {Kulkarni}, P.~{Kumar}, F.~{Lebrun}, N.~M. {Lloyd-Ronning}, C.~B.
  {Markwardt}, B.~J. {Mattson}, R.~F. {Mushotzky}, J.~P. {Norris},
  J.~{Osborne}, B.~{Paczynski}, D.~M. {Palmer}, H.~S. {Park}, A.~M. {Parsons},
  J.~{Paul}, M.~J. {Rees}, C.~S. {Reynolds}, J.~E. {Rhoads}, T.~P. {Sasseen},
  B.~E. {Schaefer}, A.~T. {Short}, A.~P. {Smale}, I.~A. {Smith}, L.~{Stella},
  G.~{Tagliaferri}, T.~{Takahashi}, M.~{Tashiro}, L.~K. {Townsley},
  J.~{Tueller}, M.~J.~L. {Turner}, M.~{Vietri}, W.~{Voges}, M.~J. {Ward},
  R.~{Willingale}, F.~M. {Zerbi}, and W.~W. {Zhang}.
\newblock {The Swift Gamma-Ray Burst Mission}.
\newblock \emph{\apj}, 611\penalty0 (2):\penalty0 1005--1020, August 2004.
\newblock \doi{10.1086/422091}.

\bibitem[{Paczynski}(1986)]{Paczynski1986}
B.~{Paczynski}.
\newblock {Gamma-ray bursters at cosmological distances}.
\newblock \emph{\apjl}, 308:\penalty0 L43--L46, September 1986.
\newblock \doi{10.1086/184740}.

\bibitem[{Blandford} and {Znajek}(1977)]{blandford1977}
R.~D. {Blandford} and R.~L. {Znajek}.
\newblock {Electromagnetic extraction of energy from Kerr black holes.}
\newblock \emph{\mnras}, 179:\penalty0 433--456, May 1977.
\newblock \doi{10.1093/mnras/179.3.433}.

\bibitem[{Kouveliotou} et~al.(1993){Kouveliotou}, {Meegan}, {Fishman}, {Bhat},
  {Briggs}, {Koshut}, {Paciesas}, and {Pendleton}]{crissa1993}
C.~{Kouveliotou}, C.~A. {Meegan}, G.~J. {Fishman}, N.~P. {Bhat}, M.~S.
  {Briggs}, T.~M. {Koshut}, W.~S. {Paciesas}, and G.~N. {Pendleton}.
\newblock {Identification of Two Classes of Gamma-Ray Bursts}.
\newblock \emph{\apjl}, 413:\penalty0 L101, August 1993.
\newblock \doi{10.1086/186969}.

\bibitem[{Janiuk} and {Yuan}(2010)]{janiuk2010}
A.~{Janiuk} and Y.~F. {Yuan}.
\newblock {The role of black hole spin and magnetic field threading the
  unstable neutrino disk in gamma ray bursts}.
\newblock \emph{\aap}, 509:\penalty0 A55, January 2010.
\newblock \doi{10.1051/0004-6361/200912725}.

\bibitem[{Thompson}(1994)]{Thompson1994}
C.~{Thompson}.
\newblock {A model of gamma-ray bursts.}
\newblock \emph{\mnras}, 270:\penalty0 480--498, October 1994.
\newblock \doi{10.1093/mnras/270.3.480}.

\bibitem[Janiuk and Sapountzis(2018)]{Janiuk2018}
Agnieszka Janiuk and Konstantinos Sapountzis.
\newblock Gamma ray bursts: Progenitors, accretion in the central engine, jet
  acceleration mechanisms.
\newblock In Zbigniew Szadkowski, editor, \emph{Cosmic Rays}, chapter~2.
  IntechOpen, Rijeka, 2018.
\newblock \doi{10.5772/intechopen.76283}.
\newblock URL \url{https://doi.org/10.5772/intechopen.76283}.

\bibitem[{Fishbone} and {Moncrief}(1976)]{Moncrief1976}
L.~G. {Fishbone} and V.~{Moncrief}.
\newblock {Relativistic fluid disks in orbit around Kerr black holes.}
\newblock \emph{\apj}, 207:\penalty0 962--976, August 1976.
\newblock \doi{10.1086/154565}.

\bibitem[{Chakrabarti}(1985)]{Chakrabarti1985}
S.~K. {Chakrabarti}.
\newblock {The natural angular momentum distribution in the study of thick
  disks around black holes}.
\newblock \emph{\apj}, 288:\penalty0 1--6, January 1985.
\newblock \doi{10.1086/162755}.

\end{thebibliography}
\end{document}